\documentclass[12pt]{iopart}

\usepackage{epsfig} 
\begin{document}

\title{Magnetic structure of CeRhIn$_{5}$ under magnetic field}

\author{S. Raymond, E. Ressouche, G. Knebel, D. Aoki and J. Flouquet}

\address{CEA-Grenoble, DRFMC, SPSMS, 38054 Grenoble Cedex 9, France}

\begin{abstract}
The magnetically ordered ground state of CeRhIn$_{5}$ at ambient pressure and zero magnetic field is an incomensurate helicoidal phase with the propagation vector $\bf{k}$=(1/2, 1/2, 0.298) and the magnetic moment in the basal plane of the tetragonal structure. We determined by neutron diffraction the two different magnetically ordered phases of CeRhIn$_{5}$ evidenced by bulk measurements under applied magnetic field in its basal plane. The low temperature high magnetic phase corresponds to a sine-wave structure of the magnetization being commensurate with $\bf{k}$=(1/2, 1/2, 1/4). At high temperature, the phase is incommensurate with $\bf{k}$=(1/2, 1/2, 0.298) and a possible small ellipticity. The propagation vector of this phase is the same as the one of the zero-field structure. 
\end{abstract}

\maketitle

The interplay between magnetism and superconductivity is one of the most important point of interest in the study of heavy fermion systems \cite{Flouquet}. The widely open question concerns the cooperative versus the antagonist nature of the two ground states. The generic phase diagram obtained in heavy fermion systems, high-T$_{c}$ superconductors and organics is composed of a superconducting region appearing in the vicinity of a vanishing magnetic phase. Because magnetic fluctuations are expected to be strong in this region of the phase diagram, they are invoked to be responsible for the Cooper pair formation \cite{Mathur}. However no definitive symbiosis between experiment and theory firmly establishes this point in the same manner than the classical  scenario of phonon mediated conventional superconductivity does. In contrast, the SO(5) theory describes the two phenomena, magnetism and superconductivity, in a unified picture via a superspin order parameter that encompases both states \cite{Demler}. This leads to a rich variety of possible phase diagrams with yet limited experimental examples to test in detail this  theory \cite{Kitaoka}. 
To this respect, CeRhIn$_{5}$ (and the related CeTIn$_{5}$ compounds with T=Ir, Co \cite{LANL}) provides a unique experimental case 
where the N\'eel ($T_{N}$) and the superconducting transition ($T_{c}$) temperatures are of the same order (of about 1 K) and can be tuned by  pressure ($p$) or magnetic field ($H$). The resulting intricated  magnetic and superconducting ($p$, $T$, $H$) phase diagram of CeRhIn$_{5}$ points toward competition between antiferromagnetism and superconductivity \cite{Park,Knebel,Chen} :  At zero magnetic field, there is a pressure range (1.6-1.9 GPa) where magnetism and superconductivity coexist with $T_{N} > T_{c}$. From 2 GPa, where $T_{N} \approx T_{c}$, a pure superconducting phase emerges and antiferromagnetism is suppressed. The application of a magnetic field in this phase induces a spectacular reentrance of the long range magnetic order. Beyond NMR experiments \cite{Kawasaki}, microscopic probes are lacking to address the nature of the magnetically ordered phases in the ($p$, $T$, $H$) phase diagram. In this viewpoint, and as a starting point for further investigations to be performed under pressure, we have investigated the magnetically ordered  phases of CeRhIn$_{5}$ under magnetic field applied in its basal plane at zero pressure. 
\begin{figure}[h]
    \centering
    \epsfig{file=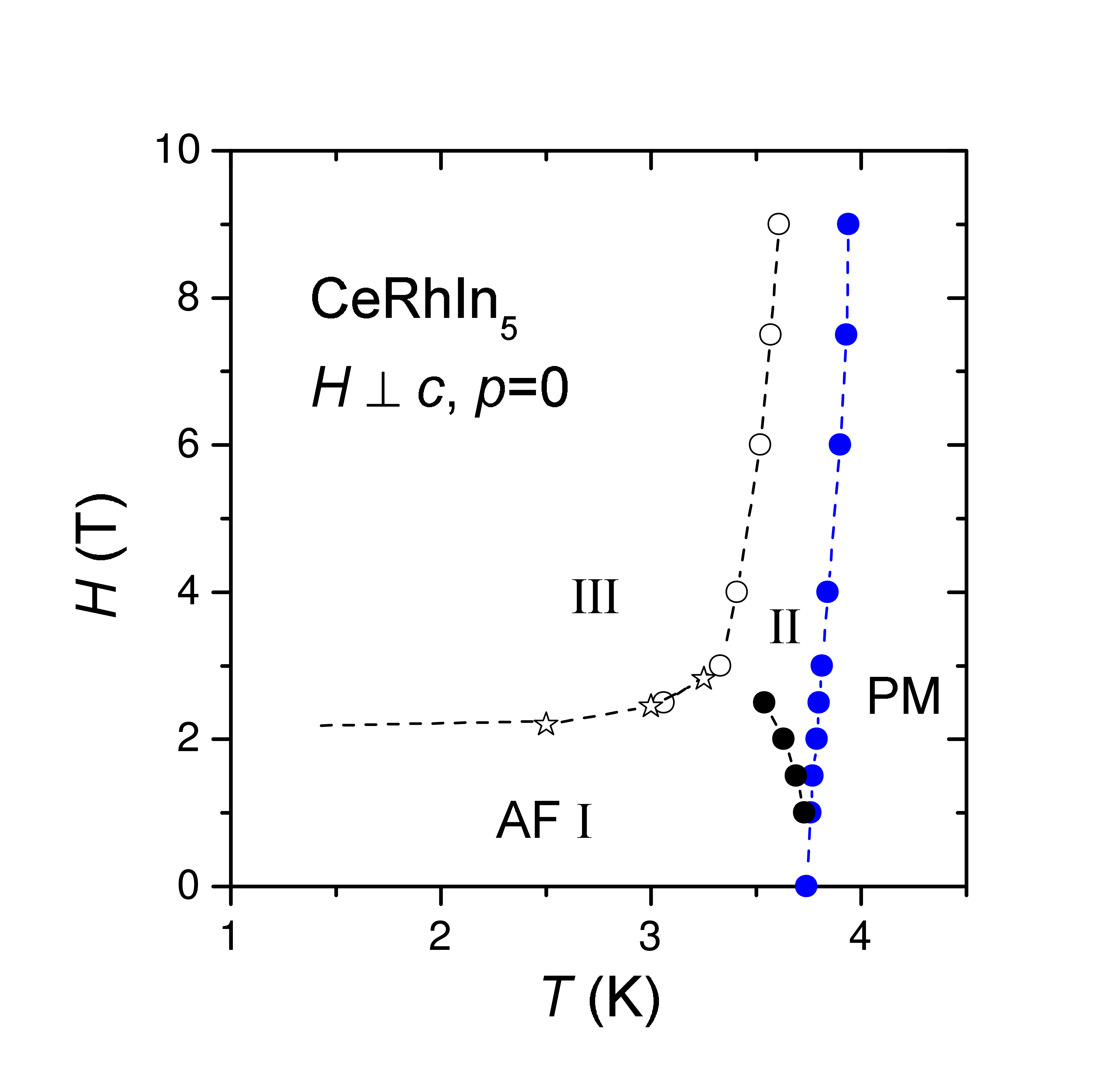,width=8cm}
    \caption{($T$, $H$) phase diagram of CeRhIn$_{5}$ determined by specific heat for field applied in the basal plane at ambient pressure. It shows three different ordered phases. Open (respect. full) symbols  correspond to first (respect. second) order transition.}
    \label{fig1}
\end{figure}

\bigskip

CeRhIn$_{5}$ crystallises in the tetragonal space group P4/mmm \cite{Moscho}. The sample was obtained by the In self flux method. A rectangular-shaped platelet of width 1 mm normal to the $c$-axis  was cut from this batch, the other dimensions being 4.3 mm along [1, -1, 0] and 2.7 mm along [1, 1, 0]. This geometry is aiming to minimize  the strong absorption cross section from In and Rh for the study of the  ([1, 1, 0], [0, 0, 1]) scattering plane. The measurements were performed on the two-axis D23-CEA-CRG (Collaborating Research Group) thermal-neutron diffractometer equipped with a lifting detector at the Institut Laue Langevin (ILL), Grenoble. A copper monochromator provides an unpolarized beam with a wavelength of $\lambda$=1.276 $\AA$. The sample was mounted in a  vertical field $^{4}$He flow cryomagnet with the [1, -1, 0] axis along the magnetic field.

The ($T$, $H$) phase diagram obtained by calorimetry measurements for the field applied perpendicular to the tetragonal axis is shown in Fig.1. It is composed of three magnetically ordered phases (two being induced by the magnetic field) consistently with the data obtained by other goup using calorimetry \cite{Cornelius}, thermal expansion and magnetostriction \cite{Correa}. In the diffraction experiment, we apply the field along [1, -1, 0] and refer to this phase diagram by neglecting the in-plane anisotropy. The magnetic structure at zero field is known to be incommensurate with slightly different propagation vectors reported in the literature, $\bf{k}$=(1/2, 1/2, 0.297) \cite{Bao} or $\bf{k}$=(1/2, 1/2, 0.298) \cite{Majumdar}. The helicoidal nature of the order,  as opposed to a sine-wave modulated structure, is known from the distribution of hyperfine field observed in NQR measurements \cite{Curro}.

In the present experiment, the lattice parameters were obtained from the centering of 18 independent reflections of the crystal and a refinement of the nuclear structure was performed at 1.9 K with 181 Bragg peaks yielding the structural parameters shown in Table 1 and the scale factor for calculation the magnetic structure. These parameters are consistent with the one of the literature \cite{Moscho} as concern the lattice parameters and the fractional coordinate $z$. 
The principal mean square atomic deplacements $u$ have typical values of such intermetallic compounds.
All refinements were corrected from extinction and absorption with the linear absorption coefficient $\mu$=0.49 mm$^{-1}$.
\Table{\label{tabl2}Structural parameters at $T$ = 1.9 K.}
\br
a = 4.638 $\AA$\\
c = 7.521  $\AA$\\
\mr
$z$&0.30526 (14)\\
$u_{Ce}$&0.0014 (5) $\AA^{2}$\\
$u_{Rh}$&0.0006 (4) $\AA^{2}$\\
$u_{In1}$&0.0018 (5) $\AA^{2}$\\
$u_{In2}$&0.0015 (4) $\AA^{2}$\\
\br
\end{tabular}
\item[] $R$ = 0.0532
\end{indented}
\end{table}
As far as magnetic scattering is concerned, the measured neutron Bragg  intensity after correction for scale factor, extinction, absorption and Lorentz factor, is the square of the component of the magnetic structure factor perpendicular to $\bf{Q}$ :
$|\mathbf{F_{M\bot}(Q)}|^{2}$. In the present case with only one magnetic Ce atom/unit cell at the origin, the magnetic structure factor is :
\begin{equation}
\mathbf{F_M(Q)}=pf(\mathbf{Q}).\mathbf{m_k}.e^{-W_{Ce}}
\end{equation}
where $p$ $\approx$ 0.27$\times$10$^{-12}$ cm is the scattering amplitude at $Q$=0 for a single magnetic moment of 1 $\mu_{B}$, $f(\bf{Q})$ is the Ce magnetic form factor, $W_{Ce}$ is the Debye-Waller factor of Ce. $\bf{m_{k}}$ is the Fourier component of the magnetic moment distribution.
The magnetic structures of interest for the present paper are (i) the collinear sine-wave structure, for which :
\begin{equation} 
\mathbf{m_{k}}=\frac{A_{k}}{2}\mathbf{u_{k}}e^{i\Phi_{k}}
\end{equation}
and (ii) the non-collinear elliptical structure :
\begin{equation}
\mathbf{m_k}=\frac{1}{2}(m^{u}\mathbf{u_k}+im^{v}\mathbf{v_k})e^{i\Phi_{k}}
\end{equation}
where $A_{k}$ is the amplitude of the sine-wave, $\bf{u_{k}}$ and $\bf{v_{k}}$ are unit vectors, $\Phi_{k}$ is a phase factor and $m^{u}$, $m^{v}$are the component of the magnetic moment along the unit vectors $\bf{u_{k}}$ and $\bf{v_{k}}$. The helicoidal order corresponds to the particular case $m^{u}$=$m^{v}$. 

The obtained propagation vector for the zero field magnetic structure is found to be $\bf{k}$=(1/2, 1/2, 0.298) in agreement with the literature. The structure was determined by measuring 16 magnetic peaks and by performing a least square fitting of the helicoidal model.  The comparison between the observed intensities and the calculated ones is shown in Table 2 with the given weighted least square factor $R$.  A magnetic moment $m_{I}$=0.59 (1) $\mu_{B}$ is found at 1.9 K, a value a little lower than the one found in the literature 0.75 (2) $\mu_{B}$ at  1.4 K \cite{Bao}. Given the rather flat temperature evolution of the order parameter between 1.4 and 1.9 K \cite{Bao}, the difference in the magnetic moment determination is not due to the difference in the measurement temperature. We believe that this difference is related to the data treatment, the present work including absorption and exctinction corrections.

\Table{\label{tabl2}Magnetic refinement with an helicoidal structure at zero field in phase I at $T$ = 1.9 K. The $\textbf{Q}$ vector is the Brillouin zone center +/- the propagation vector $\bf{k}$=(1/2, 1/2, 0.298).}
\br
\ns
\textbf{Q}& $|\mathbf{F_{M\bot}(Q)}|^{2}_{calc}$ & $|\bf{F_{M\bot}(Q)}|^{2}_{obs}$\\
\mr
(1, 1, 0) - &1.10&1.17\\
(0, 0, 0) + &1.10&1.03\\
(0, 0 ,1) +&1.52&1.50\\
(-1, -1, 1) +&1.52&1.39\\
(0, -1, 1) +&1.52&1.43\\
(1, 1, 1) -&1.30&1.36\\
(1, 1, 1) +&0.76&0.46\\
(0, 0, 2) + &1.54&1.50\\
(1, 1, 2) -&1.57&1.63\\
(0, 0, 2) -&1.57&1.53\\
(1, 0, 2) -&1.57&1.47\\
(2, 2 ,2) -&0.78&0.57\\
(1, 1, 3) -&1.47&1.62\\
(0, 0, 4) +&1.08&1.34\\
(1, 1, 4) -&1.24&1.14\\
(1, 1, 5) -&0.97&1.94\\
\br
\end{tabular}
\item[] \item[] $R$ = 0.0696
\end{indented}
\end{table}

Figure 2 show $\textbf{Q}$-scans performed along the $c$-axis for $H$ = 3 and 5 T (Phase III) with the same scan performed at $H$ = 0 T as a reference (Phase I). The propagation vector is now commensurate being (1/2, 1/2, 1/4). For this phase, 7 magnetic reflections were collected at $H$ = 3 T and $T$ = 1.9 K. The best refinement  is obtained for a colinear sine-wave structure (See Table 3) with the moment perpendicular to the field i.e. along [1, 1, 0]. Refinement with an helical structure does not work. For completeness,  an elliptic structure was refined and yields, within the error bars, zero component of the magnetic moment along the field  and thus confirms the sine-wave refinement. The propagation vector $\bf{Q}$=(1/2, 1/2, 1/4) corresponds to a particular case of the sine-wave. For a phase $\Phi_{k}$=-$\pi$/4 in eq.(2),  all the magnetic moments have the same length and the magnetic structure corresponds to the so-called ++ - - structure consisting in up, up, down, down sequence of magnetic moment when moving along the $c$-axis. This structure is favorized at low temperature because it reduces the magnetic entropy. The obtained magnetic amplitude of the sine-wave at 1.9 K is $A_{III}$=0.84 (2) $\mu_{B}$. For the peculiar ++ - - structure, the magnetic moment $m_{III}$ is related to the sine wave amplitude by $m_{III}$=$A_{III}$/$\sqrt2$. We thus obtain $m_{III}$=0.59 $\mu_{B}$, the same value than $m_{I}$. Note that the maximum in plane magnetic moment sustended by the doublet ground state is 0.92 $\mu_{B}$ as deduced from crystal field spectroscopy \cite{Christianson2}. The difference between the paramagnetic moment of the doublet ground state and the saturated ordered moment is often ascribed to the Kondo effect in cerium compounds.
\begin{figure}[h]
    \centering
    \epsfig{file=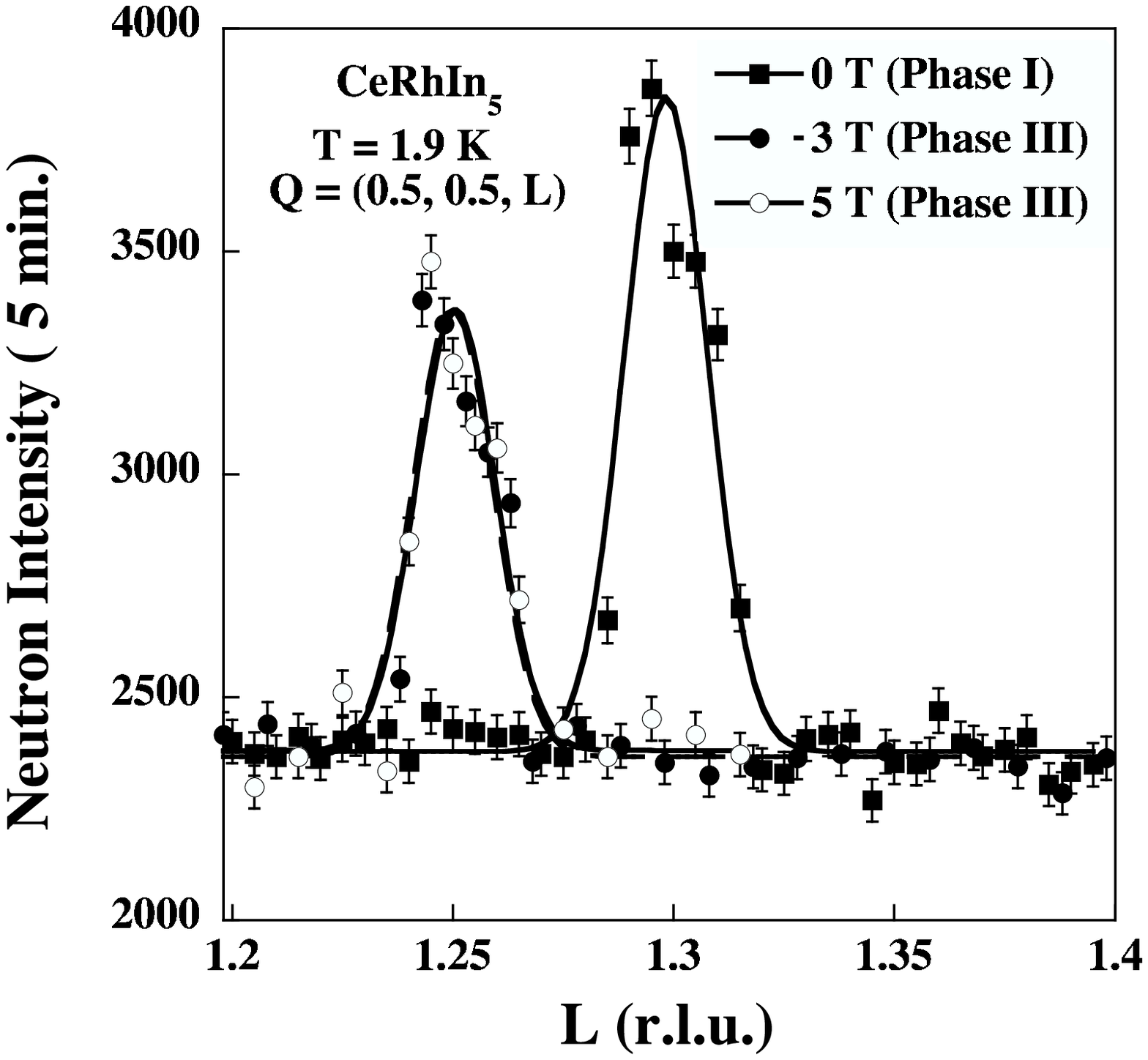,width=8cm}
    \caption{$\textbf{Q}$-scans performed along the $c$-axis for $H$= 0, 3 and 5 T at 1.9 K.}
    \label{fig2}
\end{figure}
\Table{\label{tabl2}Magnetic refinement with a sine-wave structure in phase III for $H$ = 3 T and $T$ = 1.9 K.}
\br
\ns
\bf{Q}& $|\bf{F_{M\bot}(Q)}|^{2}_{calc}$ & $|\mathbf{F_{M\bot}(Q)}|^{2}_{obs}$\\
\mr
(0, 0, 1) +&1.08&1.02\\
(1, 1, 1) -&0.62&0.65\\
(0, 0, 2) +&1.40&1.42\\
(1, 1, 2) -&1.32&1.31\\
(1, 1, 3) -&1.38&1.30\\
(0, 0, 4) +&1.08&1.00\\
(1, 1, 4) -&1.19&1.82\\
\br
\end{tabular}
\item[] $R$ = 0.0934
\end{indented}
\end{table}

Phase II was investigated by performing $\bf{Q}$-scans at 3.7 K and 4 T. An example of such a scan along the $c$-axis is shown on Fig.3a) for $\bf{Q}$=(0.5, 0.5, L)  with the same scan performed at 3.1 K in phase III as a reference. The propagation vector is found to be the same than the helicoidal phase, i.e. $\bf{k}$=(1/2, 1/2, 0.298). Figure 3b) shows the temperature variation of the magnetic Bragg peak $\bf{Q}$=(0.5, 0.5, 1.298) at 4 T. The difficulty to study this phase is that it exits in a reduced temperature range in the vicinity of the N\'eel temperature, where magnetic moment is barely developped. As a consequence the magnetic signal is weak. Figure 4 shows the field dependence of the magnetic Bragg peak intensity measured at $\bf{Q}$=(1/2, 1/2, 1.298) at 3.6 K. Since the intensity is constant in both phases, this suggests that the propagation vector does not change as a function of field. Because of the weak signal, only 4 magnetic reflections were collected in phase II at 3.6 K and 4 T and the result of a refinement with a sine-wave structure is given in Table 4. For $H$=4 T and $T$=3.6 K, the magnetic amplitude is found to be $A_{II}$=0.44 (2) $\mu_{B}$. Refinement with an elliptical phase is slightly better ($R$=0.1449 instead of $R$=0.1902) and gives a non zero component along the field $m_{[1, -1, 0]}$=0.12 (5) $\mu_{B}$, the component perpendicular to the field being then $m_{[1, 1, 0]}$=0.4 $\mu_{B}$. We cannot definitivelly conclude on the elliptical nature of this phase given the weak number of collected reflections. 
\begin{figure}[h]
    \centering
    \epsfig{file=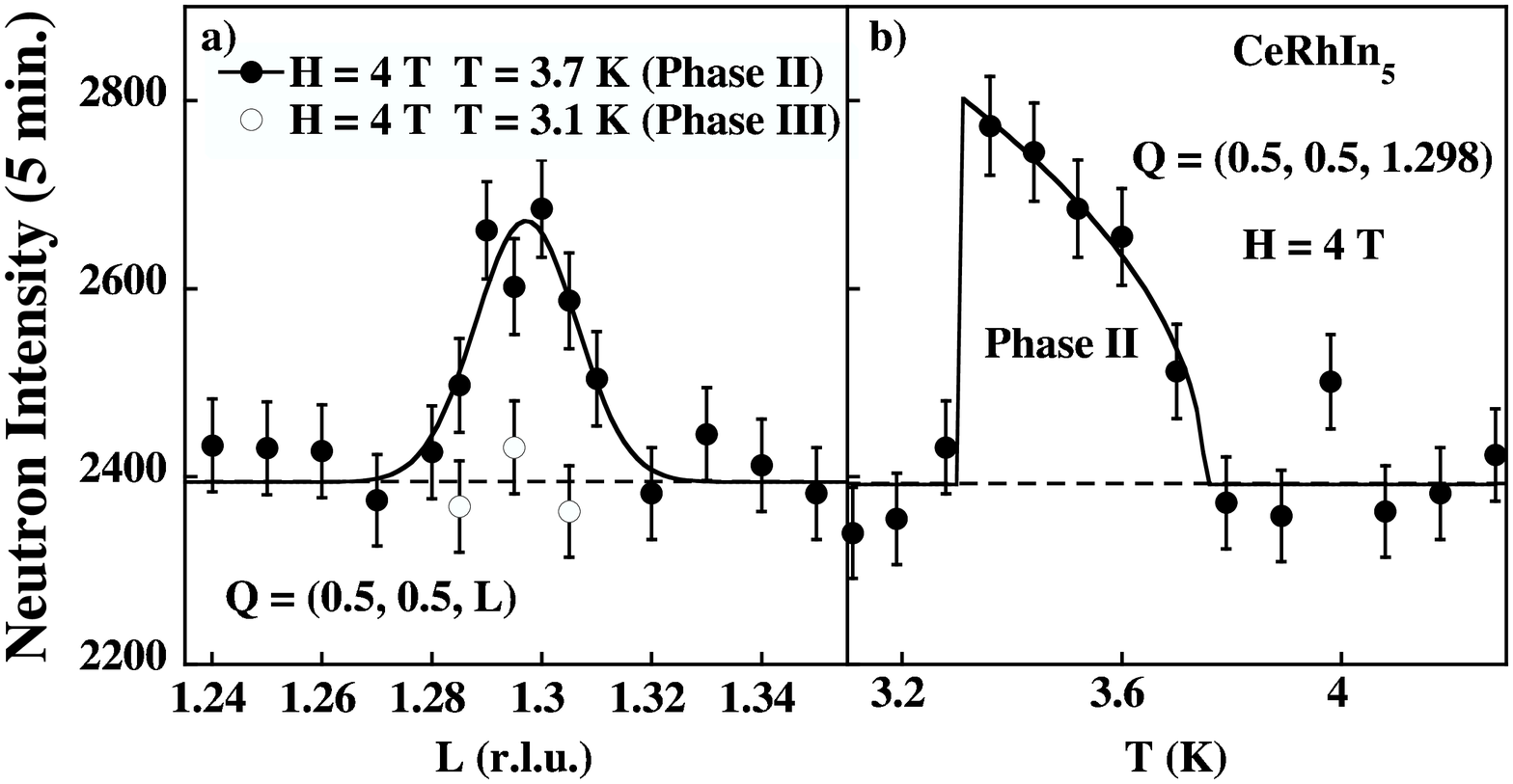,width=12cm}
    \caption{a) $\textbf{Q}$-scans performed along the $c$-axis for $H$= 4 T at  3.1 and 3.7 K. b) Temperature dependence of the Bragg peak intensity at $\textbf{Q}$=(0.5, 0.5, 1.295) for $H$ = 4 T. Solid lines are guides for the eyes. Dashed line represents the background.}
    \label{fig2}
\end{figure}

\Table{\label{tabl2}Magnetic refinement with a sine-wave structure in phase II for $H$ = 4 T and $T$ = 3.6 K.}
\br
\ns
\bf{Q}& $|\mathbf{F_{M\bot}(Q)}|^{2}_{calc}$ & $|\mathbf{F_{M\bot}(Q)}|^{2}_{obs}$\\
\mr
(1,  1, 0)-&0.04&0.07\\
(0, 0, 1)+&0.31&0.21\\
(1, 1, 1)-&0.16&0.19\\
(1, 1, 2)-&0.36&0.37\\
\br
\end{tabular}
\item[] $R$ = 0.1902
\end{indented}
\end{table}

\begin{figure}[h]
    \centering
    \epsfig{file=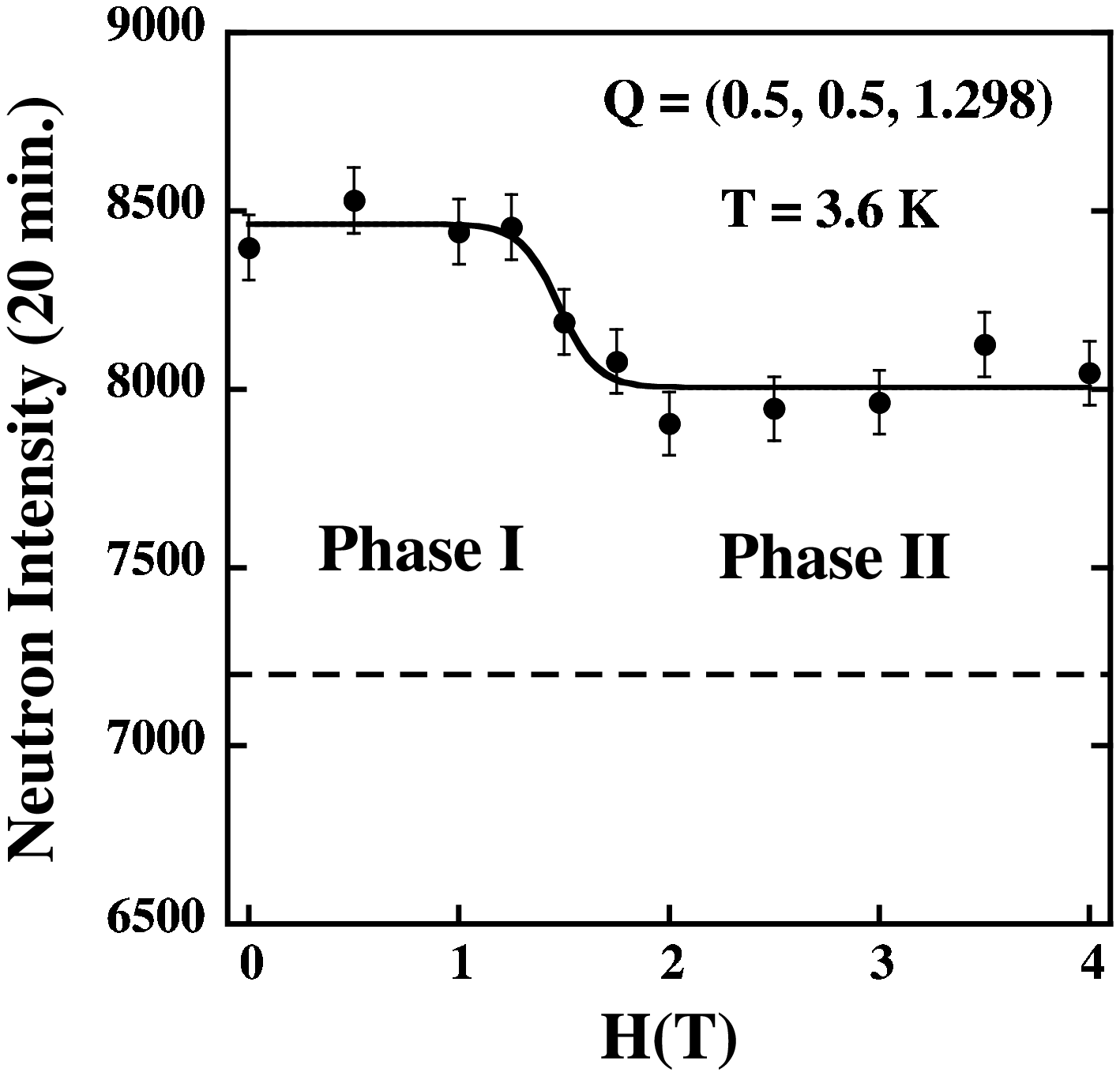,width=8cm}
    \caption{Magnetic field dependence of the Bragg peak intensity at $\bf{Q}$=(0.5, 0.5, 1.298) at 3.6 K. The solid line is a guide for the eyes. The dashed line corresponds to the background.}
    \label{fig2}
\end{figure}

\bigskip
In the previous paragraphs, we neglect the possible ferromagnetic component along the applied field. The corresponding signal was not  observed in the present experiment due to its location on the top of the nuclear peaks. The resulting structure obtained by combining the sine-wave and the ferromagnetic component  is a so-called fan structure. The fact that $m_{III}$ and $m_{I}$ are equal within the error bars indicate  that this ferromagnetic component is anyway very weak at least at low temperature. Magnetization measurements performed at 1.3 K in the basal plane give an induced ferromagnetic moment of about 0.08 $\mu_{B}$ at 5 T  \cite{Takeuchi}. The helicoidal nature of the ordering at zero field is certainly due to the RKKY interactions that allow the conditions for stabilizing such a state due to their oscillating nature. We invoke RKKY interactions rather than Fermi surface nesting because dHvA experiments suggest the localized nature of the magnetism of CeRhIn$_{5}$ at ambient pressure \cite{Onuki}. The effect of a magnetic field applied in the plane of an helix is known from a long time and was worked out shortly after the discovery of the helix structure  \cite{Nagamiya1}. The resulting sinusoidal oscillating structure or elliptical arrangement depends of the anisotropy and the magnetic field and the details of the complete ($T$, $H$) phase diagram depend on the precise Hamiltonian.  On general ground and at the mean field level, the possible transition from helix to commensurate structure under field was also predicted in the earlier works for peculiar values of the propagation vectors \cite{Nagamiya2}. A field induced transition to the antiferromagnetic state is expected for $k$ $\approx$ 1/2 and to the ++ - - structure for $k$ $\approx$ 1/4, the situation encoutered in the present work.

Despite the proximity of the zero field propagation vector to the one of the ++ - - structure, another commensurate structure is reported at zero field for CeRhIn$_{5}$ based systems with this time $\bf{k}$=(1/2, 1/2, 1/2). This antiferromagnetic order occurs in CeRh$_{1-x}$Ir$_{x}$In$_{5}$ (x) \cite{Christianson} and in CeRh$_{0.6}$Co$_{0.4}$In$_{5}$ \cite{Yokoyama}. Interestingly it is reported to coexist with the incommensurate order and also with the superconducting ground state. On cooling the incommensurate order appears first followed by the commensurate order and the superconducting state. On another hand, it is worthwhile to note that the commensurate order with $\bf{k}$=(1/2, 1/2, 1/2) alone is reported for the related CeCoIn$_{5}$ compound doped with 10 \% Cd both in the antiferromagnetic and antiferromagnetic plus superconducting phases \cite{Nicklas}. Contrastingly, the occurence of commensurate order is not reported in the diffraction studies performed on CeRhIn$_{5}$ under pressure. However  different groups obtain different results. Either  the incommensurate order is reported to change weakly with pressure up to 1.63 GPa \cite{Llobet} or at opposite, the propagation vector changes to $\bf{k}$=(1/2, 1/2, 0.396) at 0.1 GPa \cite{Majumdar}. This confusing situation asks for new experiments under pressure. The occurence of different commensurate and incommensurate phases in the ($T$, $H$, $p$, $x$) phase diagram of CeRhIn$_{5}$ deserves further investigation especially for  the interplay between magnetic order and superconductivity. 

\bigskip
We have determined the two different magnetic ordering states in CeRhIn$_{5}$ at ambient pressure under magnetic field applied in its basal plane. The low temperature phase is characterized by the commensurate propagation vector $\bf{k}$=(1/2, 1/2, 1/4) and a colinear structure with the magnetic moment perpendicular to the field. The saturated magnetic moment of 0.6 $\mu_{B}$ is the same as the one found in the zero field phase. The high temperature phase is incommensurate with the same propagation vector as the zero field incommensurate helix, $\bf{k}$=(1/2, 1/2, 0.298). The structure is colinear at first approximation with an eventual ellipticity of about 1/3.

\section*{Acknowledgements}

We acknowledge M. Zhitomirsky for illuminating discussion concerning helicoidal structures under applied magnetic field.


\section*{References}

\begin{thebibliography}{}
\bibitem{Flouquet} See e.g. J. Flouquet, Prog. Low. Temp. Phys. 2005 $\bf{15}$ Chapter 2. 
\bibitem{Mathur} See e.g. N.D. Mathur et al., Nature 1998 $\bf{394}$ 39.
\bibitem{Demler} E. Demler, W. Hanke and S.C. Zhang, Rev. Mod. Phys. 2004 $\bf{76}$ 909.
\bibitem{Kitaoka} Y. Kitaoka et al., J. Phys. Condens. Matter 2001 $\bf{13}$ L79.
\bibitem{LANL} See e.g. J.D. Thompson et al., J. Magn. Magn. Mat. 2001 $\bf{226-230}$ 5.
\bibitem{Park} T. Park et al., Nature 2006 $\bf{440}$ 65.
\bibitem{Knebel} G. Knebel et al., Phys. Rev. B 2006 $\bf{74}$ 020501(R).
\bibitem{Chen} G.F. Chen et al., Phys. Rev. Lett. 2006 $\bf{97}$ 017005. 
\bibitem{Kawasaki} S. Kawasaki et al., Phys. Rev. Lett. 2003 $\bf{91}$ 137001.
\bibitem{Moscho} E.G. Moschopoulu et al., Applied Physics A 2002 $\bf{74}$ Suppl. 5895.
\bibitem{Cornelius} A.L. Cornelius et al., Phys. Rev. B 2001 $\bf{64}$ 144111.
\bibitem{Correa} V.F. Correa et al., cond-mat/0411359.s
\bibitem{Bao} W. Bao et al., Phys. Rev. B 2000 $\bf{62}$ R14621 and Phys. Rev. B 2003 $\bf{67}$ 099903 (E).
\bibitem{Majumdar} S. Majumdar et al., Phys. Rev. B 2002 $\bf{66}$ 212502.
\bibitem{Curro} N.J. Curro et al., Phys. Rev. B 2000 $\bf{62}$ R6100.
\bibitem{Christianson2} A.D. Christianson et al., Phys. Rev. B 2002 $\bf{66}$ 193102.
\bibitem{Takeuchi} T. Takeuchi et al., J. Phys. Soc. Japan 2001 $\bf{70}$ 877.
\bibitem{Onuki} Y. Onuki et al., Acta Phys. Pol. B 2003 $\bf{34}$ 667.
\bibitem{Nagamiya1} T. Nagamiya, Solid State Phys. 1967 $\bf{20}$ 305, New York, Academic Press and references therein.
\bibitem{Nagamiya2} See e.g. Nagamiya et al., Prog. Theor. Phys. 1962 $\bf{27}$ 1253.
\bibitem{Christianson} A.D. Christianson et al., Phys. Rev. Lett. 2005 $\bf{95}$ 2117002.
\bibitem{Yokoyama} M. Yokoyama et al., J. Phys. Soc. Japan 2006 $\bf{75}$ 103703.
\bibitem{Nicklas} M. Nicklas et al., cond-mat/0703703.
\bibitem{Llobet} A. Llobet al al., Phys. Rev. B 2004 $\bf{69}$ 024403.
\end{thebibliography}
\end{document}